\documentclass[12pt]{article}
\usepackage{epsfig}
\begin{document}
\title{Study of a possibility of observation of hidden-bottom pentaquark resonances
in bottomonium photoproduction on protons and nuclei near threshold}
\author{E. Ya. Paryev$^{1,2}$ \\
{\it $^1$Institute for Nuclear Research, Russian Academy of Sciences,}\\
{\it Moscow 117312, Russia}\\
{\it $^2$Institute for Theoretical and Experimental Physics,}\\
{\it Moscow 117218, Russia}}

\renewcommand{\today}{}
\maketitle

\begin{abstract}
   We study the $\Upsilon(1S)$ meson photoproduction on protons and nuclei at the
   near-threshold center-of-mass energies below 11.4 GeV (or at the corresponding
   photon laboratory energies $E_{\gamma}$ below 68.8 GeV).
   We calculate the absolute excitation functions for the non-resonant and resonant
   photoproduction of $\Upsilon(1S)$ mesons off protons at incident photon laboratory energies of
   63--68 GeV by accounting for direct (${\gamma}p \to {\Upsilon(1S)}p$) and two-step
   (${\gamma}p \to P^+_b(11080,11125,11130)\\ \to {\Upsilon(1S)}p$) $\Upsilon(1S)$ production channels
   within different scenarios for the non-resonant total cross section of elementary reaction
   ${\gamma}p \to {\Upsilon(1S)}p$ and for branching ratios of the decays\\
   $P^+_b(11080,11125,11130) \to {\Upsilon(1S)}p$.
   We also calculate an analogous functions for photoproduction of $\Upsilon(1S)$ mesons
   on $^{12}$C and $^{208}$Pb target nuclei in the near-threshold center-of-mass beam energy region
   of 9.0--11.4 GeV by considering respective incoherent direct (${\gamma}N \to {\Upsilon(1S)}N$)
   and two-step
   (${\gamma}p \to P^+_b(11080,11125,11130) \to {\Upsilon(1S)}p$, 
   ${\gamma}n \to P^0_b(11080,11125,11130)\\ \to {\Upsilon(1S)}n$) $\Upsilon(1S)$ production processes
   within a nuclear spectral function approach.
   We show that a detailed scan of the
   $\Upsilon(1S)$ total photoproduction cross section on a proton and nuclear targets
   in the near-threshold energy region in future high-precision experiments at the proposed
   high-luminosity electron-ion colliders EIC and EicC in the U.S. and China should give a definite
   result for or against the existence of the non-strange hidden-bottom pentaquark states
   $P_{bi}^+$ and $P_{bi}^0$ ($i=$1, 2, 3) as well as clarify their decay rates.
\end{abstract}

\newpage

\section*{1. Introduction}

In a recent publication [1] the role of the new narrow hidden-charm pentaquark states $P^+_c(4312)$,
$P^+_c(4440)$ and $P^+_c(4457)$, discovered by the LHCb Collaboration in ${J/\psi}p$  invariant
mass spectrum of the $\Lambda^0_b \to K^-({J/\psi}p)$ decays [2], in near-threshold $J/\psi$ photoproduction
on nuclei has been studied in the framework of the nuclear spectral function approach by considering
both the direct non-resonant (${\gamma}N \to {J/\psi}N$) and the two-step resonant
(${\gamma}p \to P_c^+(4312)$, $P_c^+(4312) \to {J/\psi}p$;
${\gamma}p \to P_c^+(4440)$, $P_c^+(4440) \to {J/\psi}p$ and
${\gamma}p \to P_c^+(4457)$, $P_c^+(4457) \to {J/\psi}p$) $J/\psi$ elementary production processes
\footnote{$^)$It should be noted that such role of initially claimed [3] by the LHCb Collaboration pentaquark
resonance $P^+_c(4450)$ in $J/\psi$ photoproduction on nuclei at near-threshold incident photon energies
of 5--11 GeV has been investigated in our previous work [4].}$^)$
.
In the calculations the new experimental data for the total and differential
cross sections of the exclusive reaction ${\gamma}p \to {J/\psi}p$ in the threshold energy region from the GlueX
experiment [5] have been incorporated. The model-dependent upper limits on branching ratios of
$P_c^+(4312) \to {J/\psi}p$, $P_c^+(4440) \to {J/\psi}p$
and $P_c^+(4457) \to {J/\psi}p$ decays, set in this experiment, have been accounted for in them as well.

The quark structure of the above pentaquarks is $|P^+_c>=|uudc{\bar c}>$, i.e., they are composed of three
light quarks $u$, $u$, $d$ and a charm-anticharm pair $c{\bar c}$.
In a molecular scenario, due to the closeness of the observed
$P^+_c(4312)$ and $P^+_c(4440)$, $P^+_c(4457)$ masses to the ${\Sigma^+_c}{\bar D}^0$ and
${\Sigma^+_c}{\bar D}^{*0}$ thresholds, the $P^+_c(4312)$
resonance can be, in particular, considered as an S-wave ${\Sigma^+_c}{\bar D}^0$ bound state, while the
$P^+_c(4440)$ and $P^+_c(4457)$ as S-wave ${\Sigma^+_c}{\bar D}^{*0}$ bound molecular states [6--18].
The existence of molecular type hidden-charm pentaquark resonances has been predicted before the
LHCb observation [2, 3] in some earlier papers (see, for example, [19]).
It is natural to extend this picture to the bottom sector replacing the $c{\bar c}$ pair on the
bottom-antibottom $b{\bar b}$ pair as well as the non-strange $D(D^*)$ mesons on $B(B^*)$ ones and
the charmed baryons by the bottom ones. Based on the classification of hidden-charm pentaquarks
composed by a single charm baryon and $D(D^*)$ mesons, such extension has been performed in Ref. [20]
within the hadronic molecular approach. As a result, the classification of hidden-bottom pentaquarks
composed by a single bottom baryon and $B(B^*)$ mesons has been presented here. According to it,
the charged hidden-bottom partners $P^+_b(11080)$, $P^+_b(11125)$ and $P^+_b(11130)$ of the observed
hidden-charm pentaquarks $P^+_c(4312)$, $P^+_c(4440)$ and $P^+_c(4457)$, having the quark structure
$|P^+_b>=|uudb{\bar b}>$, were predicted to exist, with
masses of 11080, 11125 and 11130 MeV, respectively. Moreover, the predictions for the neutral hidden-bottom
counterparts $P^0_b(11080)$, $P^0_b(11125)$ and $P^0_b(11130)$ of the unobserved
hidden-charm states $P^0_c(4312)$, $P^0_c(4440)$ and $P^0_c(4457)$ with the quark structure
$|P^0_b>=|uddb{\bar b}>$ were provided in [20] as well.
These new exotic heavy pentaquarks can decay into the ${\Upsilon(1S)}p$ and ${\Upsilon(1S)}n$ final
states, correspondingly. They can be searched for through a scan of the cross section
\footnote{$^)$They should appear as structures at $W=11080$, 11125 and 11130 MeV or at laboratory
photon energies $E_{\gamma}=64.952$, 65.484 and 65.544 GeV in this cross section.}$^)$
of the exclusive reaction ${\gamma}p \to {\Upsilon(1S)}p$ from threshold of 10.4 GeV and up to photon
${\gamma}p$ c.m.s. energy $W=11.4$ GeV (cf. [21]).

   Therefore, it is interesting to extend the study of Ref. [1] to the consideration of
bottomonium $\Upsilon(1S)$ photoproduction on protons and nuclei near threshold to shed light on
the possibility to observe such hidden-bottom pentaquarks in this photoproduction in future
high-precision experiments at the proposed high-luminosity electron-ion colliders EIC [22--24]
and EicC [25, 26] in the U.S. and China. This is the main purpose of the present paper.
We briefly remind the main assumptions of the model [1] and describe, where it is
necessary, the corresponding extensions. We present also the predictions obtained within this
expanded model for the $\Upsilon(1S)$ excitation functions in ${\gamma}p$ as well as in
${\gamma}$$^{12}$C and ${\gamma}$$^{208}$Pb collisions at near-threshold incident energies.
They could serve as a guidance for future dedicated experiments at the above colliders.

\section*{2. The model}

\subsection*{2.1. Direct processes of non-resonant $\Upsilon(1S)$ photoproduction on nuclei}

  An incident photon can produce a $\Upsilon(1S)$ meson directly in the first inelastic ${\gamma}N$ collision.
Since we are interested in near-threshold center-of-mass photon beam energies $\sqrt{s}$ below 11.4 GeV,
corresponding to the laboratory incident photon energies $E_{\gamma}$ below 68.8 GeV or excess energies
$\epsilon_{{\Upsilon(1S)}N}$ above the ${\Upsilon(1S)}N$ threshold $\sqrt{s_{\rm th}}=m_{\Upsilon(1S)}+m_N=10.4$ GeV
($m_{\Upsilon(1S)}$ and $m_N$ are the lowest-lying bottomonium and nucleon bare masses, respectively),
$\epsilon_{{\Upsilon(1S)}N}=\sqrt{s}-\sqrt{s_{\rm th}}$ $\le$ 1.0 GeV, we have taken into account
the following direct non-resonant elementary $\Upsilon(1S)$ production processes
which have the lowest free production threshold
\footnote{$^)$We can ignore in the energy domain of our interest the contribution to the $\Upsilon(1S)$ yield
from the excited bottomonium states $\Upsilon(2S)$, $\Upsilon(3S)$ and $\chi_{\rm b}(1P)$,
$\chi_{\rm b}(2P)$ mesons feed-down due to larger their production thresholds in ${\gamma}N$ collisions.}$^)$
:
\begin{equation}
{\gamma}+p \to \Upsilon(1S)+p,
\end{equation}
\begin{equation}
{\gamma}+n \to \Upsilon(1S)+n.
\end{equation}
In what follows, in line with [27] we will neglect the modification of the outgoing $\Upsilon(1S)$
mass in nuclear matter. Also, we will ignore the medium modification of the secondary high-momentum
nucleon mass in the present work.

Disregarding the absorption of incident photons in the energy range of interest to us and describing
the $\Upsilon(1S)$ meson absorption in nuclear medium by the absorption cross section $\sigma_{{\Upsilon(1S)}N}$,
we can represent the total cross section for the production of ${\Upsilon(1S)}$ mesons
off nuclei in the direct non-resonant channels (1) and (2) of their production off target nucleons in the form [4]:
\begin{equation}
\sigma_{{\gamma}A\to {\Upsilon(1S)}X}^{({\rm dir})}(E_{\gamma})=I_{V}[A,\sigma_{{\Upsilon(1S)}N}]
\left<\sigma_{{\gamma}p \to {\Upsilon(1S)}p}(E_{\gamma})\right>_A,
\end{equation}
where
\begin{equation}
I_{V}[A,\sigma]=2{\pi}A\int\limits_{0}^{R}r_{\bot}dr_{\bot}
\int\limits_{-\sqrt{R^2-r_{\bot}^2}}^{\sqrt{R^2-r_{\bot}^2}}dz
\rho(\sqrt{r_{\bot}^2+z^2})
\end{equation}
$$
\times
\exp{\left[-A{\sigma}\int\limits_{z}^{\sqrt{R^2-r_{\bot}^2}}
\rho(\sqrt{r_{\bot}^2+x^2})dx\right]},
$$
\begin{equation}
\left<\sigma_{{\gamma}p \to {\Upsilon(1S)}p}(E_{\gamma})\right>_A=
\int\int
P_A({\bf p}_t,E)d{\bf p}_tdE
\sigma_{{\gamma}p \to {\Upsilon(1S)}p}(\sqrt{s_{\Upsilon(1S)}})
\end{equation}
and
\begin{equation}
  s_{\Upsilon(1S)}=(E_{\gamma}+E_t)^2-({\bf p}_{\gamma}+{\bf p}_t)^2,
\end{equation}
\begin{equation}
   E_t=M_A-\sqrt{(-{\bf p}_t)^2+(M_{A}-m_{N}+E)^{2}}.
\end{equation}
Here,
$\sigma_{{\gamma}p\to {\Upsilon(1S)}p}(\sqrt{s_{\Upsilon(1S)}})$ is the "in-medium"
total cross section for the production of $\Upsilon(1S)$ in reaction (1)
\footnote{$^)$In equation (3) it is supposed that the $\Upsilon(1S)$ meson production cross sections
in ${\gamma}p$ and ${\gamma}n$ interactions are the same.}$^)$
at the "in-medium" ${\gamma}p$ center-of-mass energy $\sqrt{s_{\Upsilon(1S)}}$;
$\rho({\bf r})$ and $P_A({\bf p}_t,E)$ are the local nucleon density and the nuclear
spectral function of target nucleus $A$ normalized to unity
\footnote{$^)$The concrete information about these quantities, used in our subsequent calculations,
is given in [28--30].}$^)$;
${\bf p}_{t}$  and $E$ are the internal momentum and binding energy of the struck target nucleon
just before the collision; $A$ is the number of nucleons in
the target nucleus, $M_{A}$  and $R$ are its mass and radius;
${\bf p}_{\gamma}$ and $E_{\gamma}$ are the laboratory momentum and energy of the initial photon.
Motivated by the fact that the nuclear medium suppresses $\Upsilon(1S)$ production as much as
$J/\psi$ production, we will employ for the $\Upsilon(1S)$--nucleon absorption cross section
$\sigma_{{\Upsilon(1S)}N}$ in our calculations the same value of 3.5 mb as was adopted in Ref. [4]
for the $J/\psi$--nucleon absorption cross section $\sigma_{{J/\psi}N}$ (cf. [31--33]).

  As earlier in [4], we suggest that the "in-medium" cross section
$\sigma_{{\gamma}p \to {\Upsilon(1S)}p}(\sqrt{s_{\Upsilon(1S)}})$ for $\Upsilon(1S)$ production in process (1)
is equivalent to the vacuum cross section $\sigma_{{\gamma}p \to {\Upsilon(1S)}p}({\sqrt{s}})$ in which
the vacuum center-of-mass energy squared s, presented by the formula
\begin{equation}
  s=W^2=(E_{\gamma}+m_N)^2-{\bf p}_{\gamma}^2,
\end{equation}
is replaced by the in-medium expression (6). The latter cross section has been determined experimentally
both earlier [34--36] and recently [37, 38] only at high photon--proton center-of-mass energies
$W=\sqrt{s} > 60$ GeV (see Fig. 2 given below). And up to now, the experimental data on $\Upsilon(1S)$
production in the channel ${\gamma}p \to {\Upsilon(1S)}p$ are not available in the threshold energy region
$W \le 11.4$ GeV, where the masses of the predicted [20] $P_b$ states are concentrated
and where they can be observed [21] in ${\gamma}p$ reactions.

The total cross section of this channel can be evaluated using the following indirect route. An analysis
of the data on the production of $\Upsilon(1S)$ and $J/\psi$ mesons in ${\gamma}p$ collisions in the
kinematic range of $80 < W < 160$ GeV, taken by the ZEUS Collaboration at HERA [34], gave the following
ratio of the $\Upsilon(1S)$ to $J/\psi$ photoproduction cross sections in this range:
\begin{equation}
 \sigma_{{\gamma}p \to {\Upsilon(1S)}p}(W)/\sigma_{{\gamma}p \to {J/\psi}p}(W) \approx 5\cdot10^{-3}.
\end{equation}
Accounting for the commonality in the $J/\psi$ and $\Upsilon(1S)$ production in ${\gamma}p$ interactions
[39], we assume that in the threshold region $W \le 11.4$ GeV the ratio of the total cross
sections of the reactions ${\gamma}p \to {\Upsilon(1S)}p$ and ${\gamma}p \to {J/\psi}p$
is the same as that of Eq. (9) derived at the same high ${\gamma}p$ c.m.s. energies. But now, in this
ratio the former and latter cross sections are calculated,
respectively, at the collisional energies $\sqrt{s}$ and $\sqrt{{\tilde s}}$ which correspond to the
same excess energies $\epsilon_{{\Upsilon(1S)}N}$ and $\epsilon_{{J/\psi}N}$ above the
${\Upsilon(1S)}N$  and ${J/\psi}N$ thresholds, viz.:
\begin{equation}
 \sigma_{{\gamma}p \to {\Upsilon(1S)}p}(\sqrt{s})/
 \sigma_{{\gamma}p \to {J/\psi}p}(\sqrt{{\tilde s}}) \approx 5\cdot10^{-3},
\end{equation}
where, according to above-mentioned, the center-of-mass energies $\sqrt{s}$ and $\sqrt{{\tilde s}}$
are linked by the relation:
\begin{equation}
\epsilon_{{J/\psi}N}=\sqrt{{\tilde s}}-\sqrt{{\tilde s}_{\rm th}}=
\epsilon_{{\Upsilon(1S)}N}=\sqrt{s}-\sqrt{s_{\rm th}}.
\end{equation}
Here, $\sqrt{{\tilde s}_{\rm th}}=m_{J/\psi}+m_N$ ($m_{J/\psi}$ is the bare $J/\psi$ meson mass).
With this, we have:
\begin{equation}
\sqrt{{\tilde s}}=\sqrt{s}-\sqrt{s_{\rm th}}+\sqrt{{\tilde s}_{\rm th}}=\sqrt{s}-m_{\Upsilon(1S)}+m_{J/\psi}.
\end{equation}
Evidently, that at high energies such that $\sqrt{s} >> \sqrt{s_{\rm th}}$,
$\sqrt{{\tilde s}}$ $\approx$ $\sqrt{s}$ and the expression (10) transforms to (9). At low incident photon
energies $\sqrt{s} \le 11.4$ GeV of interest, the c.m.s. energy $\sqrt{{\tilde s}} \le 5.04$ GeV. The latter
corresponds, as is easy to see, to the laboratory photon energy domain $\le$ 13.05 GeV.
For the free total cross section $\sigma_{{\gamma}p \to {J/\psi}p}({\sqrt{{\tilde s}}})$
in this domain we have adopted the following expression [1],
based on the predictions of the two gluon and three gluon exchange model [40] near threshold:
\begin{equation}
\sigma_{{\gamma}p \to {J/\psi}p}({\sqrt{{\tilde s}}})= \sigma_{2g}({\sqrt{{\tilde s}}})+
\sigma_{3g}({\sqrt{{\tilde s}}}),
\end{equation}
where
\begin{equation}
\sigma_{2g}({\sqrt{{\tilde s}}})=a_{2g}(1-x)^2\left[\frac{{\rm e}^{bt^+}-{\rm e}^{bt^-}}{b}\right],
\end{equation}
\begin{equation}
\sigma_{3g}({\sqrt{{\tilde s}}})=a_{3g}(1-x)^0\left[\frac{{\rm e}^{bt^+}-{\rm e}^{bt^-}}{b}\right]
\end{equation}
and
\begin{equation}
  x=({\tilde s}_{\rm th}-m^2_N)/({\tilde s}-m^2_N).
\end{equation}
\begin{figure}[htb]
\begin{center}
\includegraphics[width=16.0cm]{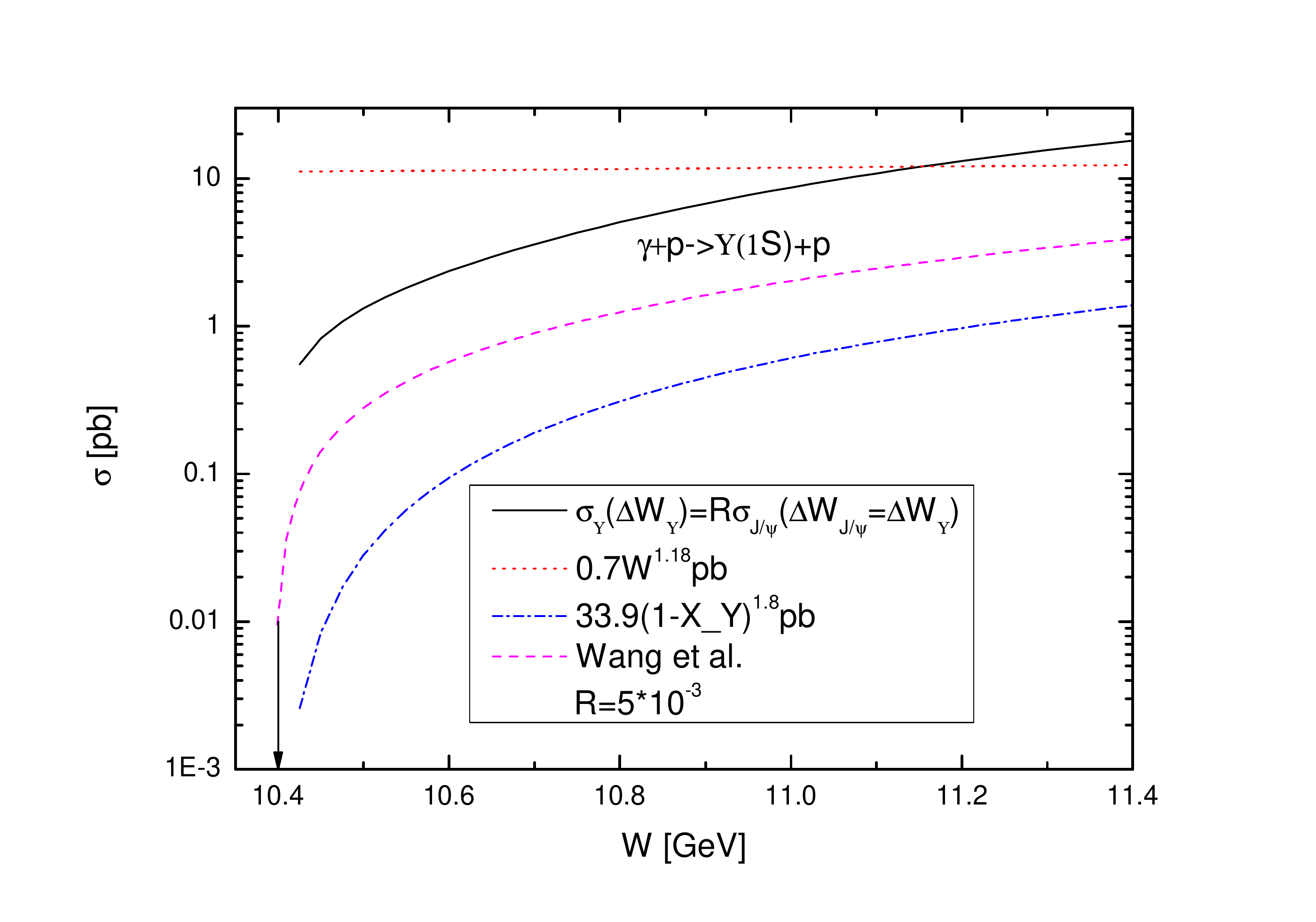}
\vspace*{-2mm} \caption{(Color online) The non-resonant total cross section for the reaction
${\gamma}p \to {\Upsilon(1S)}p$ as a function of the center-of-mass energy $W=\sqrt{s}$
of the photon--proton collisions.
Solid, dashed, dotted-dashed and dotted curves are calculations by (10)--(20),
within the dipole Pomeron model [21], by (22) and (24), respectively.
The arrow indicates the center-of-mass threshold energy
for direct $\Upsilon(1S)$ photoproduction on a free target proton being at rest.}
\label{void}
\end{center}
\end{figure}
Here, $t^+$ and $t^-$ are, respectively, the maximal and minimal values of the squared
four-momentum transfer $t$ between the incident photon and the outgoing $J/\psi$ meson.
These values correspond to the $t$ where the
$J/\psi$ is produced at angles of 0$^{\circ}$ and 180$^{\circ}$ in ${\gamma}p$ c.m.s., respectively.
They can readily be expressed in terms of the total energies and momenta of the initial photon and the $J/\psi$ meson,
$E^*_{\gamma}, p^*_{\gamma}$ and $E^*_{J/\psi}, p^*_{J/\psi}$, in this system as follows:
\begin{equation}
t^{\pm}=m_{J/\psi}^2-2E^*_{\gamma}(m_N^2)E^*_{J/\psi}(m_{J/\psi}){\pm}2p^*_{\gamma}(m_N^2)p^*_{J/\psi}(m_{J/\psi}),
\end{equation}
where
\begin{equation}
p_{\gamma}^*(m_{N}^2)=\frac{1}{2\sqrt{{\tilde s}}}\lambda({\tilde s},0,m_{N}^2),
\end{equation}
\begin{equation}
p^*_{J/\psi}(m_{J/\psi})=\frac{1}{2\sqrt{{\tilde s}}}\lambda({\tilde s},m_{J/\psi}^{2},m_N^2)
\end{equation}
and
\begin{equation}
E^*_{\gamma}(m_N^2)=p^*_{\gamma}(m_N^2), \,\,\,\,
E^*_{J/\psi}(m_{J/\psi})=\sqrt{m^2_{J/\psi}+[p^*_{J/\psi}(m_{J/\psi})]^2};
\end{equation}
\begin{equation}
\lambda(x,y,z)=\sqrt{{\left[x-({\sqrt{y}}+{\sqrt{z}})^2\right]}{\left[x-
({\sqrt{y}}-{\sqrt{z}})^2\right]}}.
\end{equation}
Parameter $b$ in Eqs. (14), (15) is an exponential $t$-slope of the differential cross
section of the reaction ${\gamma}p \to {J/\psi}p$ near threshold [40].
According to [5], $b\approx$1.67 GeV$^{-2}$. We will
employ this value in our calculations. The normalization coefficients $a_{2g}$ and $a_{3g}$ was determined
in [1] as $a_{2g}=44.780$ nb/GeV$^2$ and $a_{3g}=2.816$ nb/GeV$^2$
assuming that incoherent sum (13) saturates the total experimental cross section of the
reaction ${\gamma}p \to {J/\psi}p$ measured at GlueX [5] at photon energies around 8.38 and 11.62 GeV.
\begin{figure}[htb]
\begin{center}
\includegraphics[width=16.0cm]{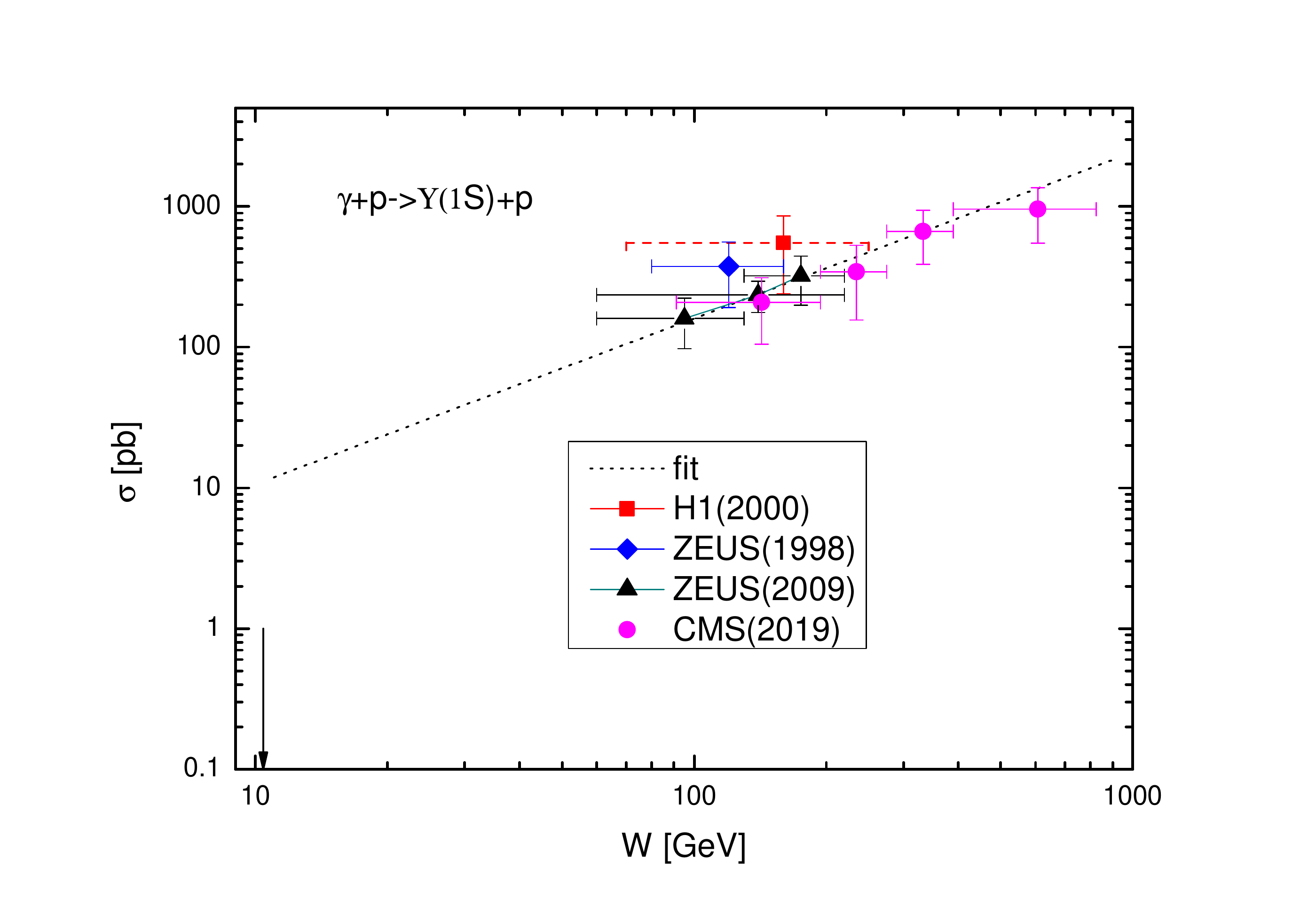}
\vspace*{-2mm} \caption{(Color online) The non-resonant total cross section for the reaction
${\gamma}p \to {\Upsilon(1S)}p$ as a function of the center-of-mass energy $W=\sqrt{s}$
of the photon--proton collisions.
Dotted curve is calculation by (24). The experimental data are from Refs. [34--37].
The arrow indicates the center-of-mass threshold energy
for direct $\Upsilon(1S)$ photoproduction on a free target proton being at rest.}
\label{void}
\end{center}
\end{figure}

  The results of calculations by Eqs. (10)--(20) of the non-resonant total cross section of the reaction
${\gamma}p \to {\Upsilon(1S)}p$ at "low" energies are shown in Fig. 1 (solid curve). In this figure we also
show the predictions from the dipole Pomeron model [21] (dashed curve)
\footnote{$^)$The author thanks X.-Y. Wang for sending these predictions to him.}$^)$
and from
recently proposed parametrization [39]
\begin{equation}
 \sigma_{{\gamma}p \to {\Upsilon(1S)}p}(\sqrt{s})=33.9(1-x_{\Upsilon})^{1.8}~[\rm pb],
\end{equation}
where $x_{\Upsilon}$ is defined as
\begin{equation}
  x_{\Upsilon}=(s_{\rm th}-m^2_N)/(s-m^2_N)
\end{equation}
(dotted-dashed curve). The results from the extrapolation of the fit [41]
\begin{equation}
 \sigma_{{\gamma}p \to {\Upsilon(1S)}p}(\sqrt{s})=0.7(\sqrt{s})^{1.18}~[\rm pb]
\end{equation}
of the high-energy data [36] (see Fig. 2
\footnote{$^)$Where also the data from other high-energy experiments [34, 35, 37] are given.}$^)$
)
to threshold energies of interest are shown in Fig. 1
as well (dotted curve).
It is, in particular, seen that at photon energies around 11 GeV our parametrization (10)--(20) is close
to the results from the high-energy fit (24), and is considerably larger (by factors of about 5 and 15,
respectively) than the results from the dipole Pomeron model [21] and from the parametrization (22).
Therefore, the use of two parametrizations (10)--(20) and (22) in our subsequent calculations
will give us the reasonable bounds for the elastic background under the pentaquark peaks.
When they are employed in the calculations of the non-resonant $\Upsilon(1S)$
photoproduction off nuclei presented below then, in line with above-mentioned,
instead of the vacuum quantity $s$, appearing in Eqs. (10)--(12) and (23), one needs to adopt
its in-medium expression (6) in which the laboratory incident photon energy $E_{\gamma}$ is expressed
via the given free space center-of-mass energy $W$ as $E_{\gamma}=(W^2-m_N^2)/(2m_N)$.
And instead of the quantity $m_N^2$, entering into Eq. (18), we should employ the difference $E_t^2-p_t^2$.

\subsection*{2.2. Two-step processes of resonant $\Upsilon(1S)$ photoproduction on nuclei}

  At photon center-of-mass energies $\le$ 11.4 GeV, an incident photons can produce a non-strange charged
$P^+_b(11080)$, $P^+_b(11125)$, $P^+_b(11130)$ and neutral $P^0_b(11080)$, $P^0_b(11125)$,
$P^0_b(11130)$ resonances with pole masses $M_{b1}=11080$ MeV, $M_{b2}=11125$ MeV, $M_{b3}=11130$ MeV,
respectively, predicted in Ref. [20] on the basis of the observed [2] three $P_c^+$ states,
in the first inelastic collisions with an intranuclear protons and neutrons
\footnote{$^)$We recall that the threshold (resonant) energies $E^{\rm R1}_{\gamma}$,
$E^{\rm R2}_{\gamma}$, $E^{\rm R3}_{\gamma}$ for the photoproduction
of $P^+_b(11080)$, $P^+_b(11125)$, $P^+_b(11130)$ and $P^0_b(11080)$, $P^0_b(11125)$, $P^0_b(11130)$
resonances
on a free target protons and neutrons being at rest are $E^{\rm R1}_{\gamma}=64.952$ GeV,
$E^{\rm R2}_{\gamma}=65.484$ GeV, $E^{\rm R3}_{\gamma}=65.544$ GeV and $E^{\rm R1}_{\gamma}=64.863$ GeV,
$E^{\rm R2}_{\gamma}=65.395$ GeV, $E^{\rm R3}_{\gamma}=65.454$ GeV, respectively.}$^)$
:
\begin{eqnarray}
{\gamma}+p \to P^+_b(11080),\nonumber\\
{\gamma}+p \to P^+_b(11125),\nonumber\\
{\gamma}+p \to P^+_b(11130);
\end{eqnarray}
\begin{eqnarray}
{\gamma}+n \to P^0_b(11080),\nonumber\\
{\gamma}+n \to P^0_b(11125),\nonumber\\
{\gamma}+n \to P^0_b(11130).
\end{eqnarray}
Further, the produced intermediate pentaquarks can decay into the final states $\Upsilon(1S)$$p$
and $\Upsilon(1S)$$n$:
\begin{eqnarray}
P^+_b(11080) \to \Upsilon(1S)+p,\nonumber\\
P^+_b(11125) \to \Upsilon(1S)+p,\nonumber\\
P^+_b(11130) \to \Upsilon(1S)+p;
\end{eqnarray}
\begin{eqnarray}
P^0_b(11080) \to \Upsilon(1S)+n,\nonumber\\
P^0_b(11125) \to \Upsilon(1S)+n,\nonumber\\
P^0_b(11130) \to \Upsilon(1S)+n.
\end{eqnarray}
Since the $P^+_{bi}$ and $P^0_{bi}$ states are not observed experimentally up to now, presently,
neither their total decay widths $\Gamma_{bi}$,
the branching ratios $Br[P^+_{bi} \to {\Upsilon(1S)}p]$ and $Br[P^0_{bi} \to {\Upsilon(1S)}n]$
\footnote{$^)$Here, $i=$1, 2, 3. $P^+_{b1}$, $P^+_{b2}$, $P^+_{b3}$ and
$P^0_{b1}$, $P^0_{b2}$, $P^0_{b3}$ stand for $P^+_b(11080)$, $P^+_b(11125)$, $P^+_b(11130)$
and  $P^0_b(11080)$, $P^0_b(11125)$, $P^0_b(11130)$, respectively.}$^)$
of decays (27) and (28) nor spin-parity quantum numbers are known in a model-independent way.
Therefore, to estimate the $\Upsilon(1S)$ production cross section from production/decay chains (25)--(28)
one needs to rely on the theoretical predictions as well as on the similarity of the basic features
of the decay properties of the $qqqb{\bar b}$ and $qqqc{\bar c}$ systems (with $q=u$ or $d$). Thus, the results
for the decay rates of the modes (27), (28) are expressed in Ref. [20] in terms of the model parameter
$\Lambda$, which should be constrained from the future experiments. The existence of the hidden-bottom pentaquark resonances with masses around 11 GeV and total decay widths from a few MeV to 45 MeV has been also predicted in
Refs. [42--44]. With this and with above-mentioned, it is natural to assume, analogously to [41],
for $P^+_{bi}$ and $P^0_{bi}$ states the same total widths $\Gamma_{bi}$ as for their hidden-charm partners
$P^+_c(4312)$, $P^+_c(4440)$ and $P^+_c(4457)$, i.e.,
$\Gamma_{b1}=9.8$ MeV, $\Gamma_{b2}=20.6$ MeV, $\Gamma_{b3}=6.4$ MeV [2]. And for all branching ratios
$Br[P^+_{bi} \to {\Upsilon(1S)}p]$ and $Br[P^0_{bi} \to {\Upsilon(1S)}n]$ of the decays (27) and (28) to adopt
in our study the same [41] three main options: $Br[P^+_{bi} \to {\Upsilon(1S)}p]=1$, 2 and 3\% and
$Br[P^0_{bi} \to {\Upsilon(1S)}n]=1$, 2 and 3\% as those used in Ref. [1] for the
$P^+_{ci} \to {J/\psi}p$ decays. In order to see additionally the size of the impact of branching
fractions $Br[P^+_{bi} \to {\Upsilon(1S)}p]$ and $Br[P^0_{bi} \to {\Upsilon(1S)}n]$
on the resonant $\Upsilon(1S)$ yield in ${\gamma}$$p$ $\to {\Upsilon(1S)}p$,
${\gamma}$$^{12}$C $\to {\Upsilon(1S)}X$ and
${\gamma}$$^{208}$Pb $\to {\Upsilon(1S)}X$ reactions, we will also calculate this yield
supposing that all these branching fractions are equal to 5 and 10\% as well.

According to [1], majority of the $P^+_{bi}$ and $P^0_{bi}$ ($i=$1, 2, 3)
resonances, having vacuum total decay widths
in their rest frames $\Gamma_{b1}=9.8$ MeV, $\Gamma_{b2}=20.6$ MeV, $\Gamma_{b3}=6.4$ MeV, respectively,
decay to $\Upsilon(1S)p$ and ${\Upsilon(1S)n}$ out of the target nuclei of interest.
As in [1] for $P_{ci}^+$ states, their free spectral functions are assumed to be described by
the non-relativistic Breit-Wigner distributions:
\begin{equation}
S_{bi}^+(\sqrt{s},\Gamma_{bi})=S_{bi}^0(\sqrt{s},\Gamma_{bi})
=\frac{1}{2\pi}\frac{\Gamma_{bi}}{(\sqrt{s}-M_{bi})^2+{\Gamma}_{bi}^2/4},\,\,\,
i=1, 2, 3;
\end{equation}
where $\sqrt{s}$ is the total ${\gamma}N$ c.m.s. energy defined above by Eq.~(8).
It should be pointed out that in case of calculating the excitation functions for production of
$P^+_{bi}$ and $P^0_{bi}$ ($i=$1, 2, 3) resonances in reactions (25) and (26)
on $^{12}$C and $^{208}$Pb targets in the "free" $P^+_{bi}$ and $P^0_{bi}$
spectral function scenario (see Fig.~4 below), this energy should be taken in form of Eq.~(6).
Spectral functions $S_{bi}^+$ and $S_{bi}^0$ correspond to the $P^+_{bi}$ and $P^0_{bi}$, respectively.
In line with [1], we assume that the in-medium spectral functions $S_{bi}^+(\sqrt{s},\Gamma_{\rm med}^{bi})$,
and $S_{bi}^0(\sqrt{s},\Gamma_{\rm med}^{bi})$
are also described by the Breit-Wigner formula (29) with a total in-medium
widths $\Gamma_{\rm med}^{bi}$ ($i=$1, 2, 3)
in their rest frames, obtained as a sum of the vacuum decay widths, $\Gamma_{bi}$, and
averaged over local nucleon density
$\rho_N({\bf r})$ collisional widths $<\Gamma_{{\rm coll},bi}>$ appearing
due to $P^+_{bi}N$, $P^0_{bi}N$ inelastic collisions:
\begin{equation}
\Gamma_{\rm med}^{bi}=\Gamma_{bi}+<\Gamma_{{\rm coll},bi}>, \,\,\,i=1, 2, 3.
\end{equation}
According to [4], the average collisional width $<\Gamma_{{\rm coll},bi}>$ has a form:
\begin{equation}
<\Gamma_{{\rm coll},bi}>={\gamma_c}{v_c}{\sigma_{P_{bi}N}}<\rho_N>.
\end{equation}
Here, $\sigma_{P_{bi}N}$ is the $P^+_{bi}$, $P^0_{bi}$--nucleon inelastic cross section
\footnote{$^)$Taking into account the quark contents of the hidden-charm and hidden-bottom pentaquarks
as well as the fact that the nuclear medium suppresses $\Upsilon(1S)$ production as much as
$J/\psi$ production, we will employ in the following calculations
for the absorption cross section $\sigma_{P_{bi}N}$ for each $P^+_{bi}$ and $P^0_{bi}$ ($i=$1, 2, 3)
the same value of 33.5 mb as was adopted in Ref. [1]
for the $P^+_{ci}$--nucleon absorption cross section.}$^)$
and the Lorentz $\gamma$-factor
$\gamma_c$ and the velocity $v_c$ of the resonances $P_{bi}^+$, $P_{bi}^0$
in the nuclear rest frame are determined by:
\begin{equation}
\gamma_c=\frac{(E_{\gamma}+E_t)}{\sqrt{s}},\,\,\,\,\,v_c=\frac{|{\bf p}_{\gamma}+{\bf p}_t|}{(E_{\gamma}+E_t)}.
\end{equation}

Within the hadronic molecular scenario of $P^+_{bi}$ and $P^0_{bi}$ states
\footnote{$^)$In this scenario, due to the proximity of the predicted
$P^+_{b1}$, $P^0_{b1}$ and $P^+_{b2}$, $P^0_{b2}$, $P^+_{b3}$, $P^0_{b3}$ masses
to the ${\Sigma_b}{B}$ and ${\Sigma_b}{B}^{*}$ thresholds [20], the $P^+_{b1}$, $P^0_{b1}$
resonances can be considered as the ${\Sigma_b}{B}$ bound states, while the
$P^+_{b2}$, $P^0_{b2}$ and $P^+_{b3}$, $P^0_{b3}$ as ${\Sigma_b}{B}^{*}$
bound molecular systems [20, 41--47].}$^)$
,
in which their spins-parities are $J^P=(1/2)^-$ for $P^+_{b1}$ and $P^0_{b1}$,
$J^P=(1/2)^-$ for $P^+_{b2}$ and $P^0_{b2}$,
$J^P=(3/2)^-$ for $P^+_{b3}$ and $P^0_{b3}$ [20, 21],
the free Breit-Wigner total cross sections for their production
in reactions (25), (26) can be described on the basis of the spectral functions (29) and known branching
fractions $Br[P^+_{bi} \to {\gamma}p]$ and $Br[P^0_{bi} \to {\gamma}n]$ ($i=$1, 2, 3)
as follows [41, 48]:
\begin{eqnarray}
\sigma_{{\gamma}p \to P^+_{bi}}(\sqrt{s},\Gamma_{bi})=f_{bi}\left(\frac{\pi}{p^*_{\gamma}}\right)^2
Br[P^+_{bi} \to {\gamma}p]S_{bi}^+(\sqrt{s},\Gamma_{bi})\Gamma_{bi},\nonumber\\
\sigma_{{\gamma}n \to P^0_{bi}}(\sqrt{s},\Gamma_{bi})=f_{bi}\left(\frac{\pi}{p^*_{\gamma}}\right)^2
Br[P^0_{bi} \to {\gamma}n]S_{bi}^0(\sqrt{s},\Gamma_{bi})\Gamma_{bi}.
\end{eqnarray}
Here, the center-of-mass 3-momentum in the incoming ${\gamma}N$ channel, $p^*_{\gamma}$,
is defined above by Eq. (18) in which one has to make the substitution ${\tilde s} \to s$
and the ratios of spin factors $f_{b1}=1$, $f_{b2}=1$, $f_{b3}=2$.

In line with [1, 41, 49], we assume that the $P^+_{b1}$ and $P^0_{b1}$ $(1/2)^-$,
$P^+_{b2}$ and $P^0_{b2}$ $(1/2)^-$,  and
$P^+_{b3}$ and $P^0_{b3}$ $(3/2)^-$ decays to ${\Upsilon(1S)}p$ and ${\Upsilon(1S)}n$
are dominated by the lowest
partial waves with relative orbital angular momentum $L=0$. Then, the branching
fractions $Br[P^+_{bi} \to {\gamma}p]$ and $Br[P^0_{bi} \to {\gamma}n]$
can be expressed, adopting the vector-meson dominance model,
respectively, through the branching ratios $Br[P^+_{bi} \to {\Upsilon(1S)}p]$
and $Br[P^0_{bi} \to {\Upsilon(1S)}n]$ in the following manner [1, 41, 48, 49]:
\begin{eqnarray}
Br[P^+_{bi} \to {\gamma}p]=4{\pi}{\alpha}\left(\frac{f_{\Upsilon}}{m_{\Upsilon(1S)}}\right)^2f_{0,bi}
\left(\frac{p^*_{\gamma,bi}}{p^*_{\Upsilon,bi}}\right)
Br[P^+_{bi} \to {\Upsilon(1S)}p],\nonumber\\
Br[P^0_{bi} \to {\gamma}n]=4{\pi}{\alpha}\left(\frac{f_{\Upsilon}}{m_{\Upsilon(1S)}}\right)^2f_{0,bi}
\left(\frac{p^*_{\gamma,bi}}{p^*_{\Upsilon,bi}}\right)
Br[P^0_{bi} \to {\Upsilon(1S)}n],
\end{eqnarray}
where $f_{\Upsilon}=$ 238 MeV [41] is the $\Upsilon(1S)$ decay constant, $\alpha=$1/137 is the electromagnetic
fine structure constant and
\begin{equation}
p_{\gamma,bi}^*=\frac{1}{2M_{bi}}\lambda(M_{bi}^2,0,m_{N}^2),\,\,\,
p^*_{\Upsilon,bi}=\frac{1}{2M_{bi}}\lambda(M_{bi}^2,m_{\Upsilon(1S)}^{2},m_N^2),\,\,\,
\end{equation}
\begin{equation}
f_{0,bi}=\frac{2}{2+{\gamma}^2_{bi}},\,\,\,\,\,{\gamma}^2_{bi}=1+p^{*2}_{\Upsilon,bi}/m^2_{\Upsilon(1S)}.
\end{equation}
Accounting for that $Br[P^+_{bi} \to {\Upsilon(1S)}p]=Br[P^0_{bi} \to {\Upsilon(1S)}n]$ [20], we obtain
from Eqs. (34)--(36) that
\begin{equation}
Br[P^0_{bi} \to {\gamma}n]=Br[P^+_{bi} \to {\gamma}p].
\end{equation}
With Eq. (33) and with this, we have:
\begin{equation}
\sigma_{{\gamma}p \to P^+_{bi}}(\sqrt{s},\Gamma_{bi})=\sigma_{{\gamma}n \to P^0_{bi}}(\sqrt{s},\Gamma_{bi}).
\end{equation}
Eqs. (35), (36) yield that ($p_{\gamma,b1}^*,p^*_{\Upsilon,b1},f_{0,b1})=$(5.500~GeV/c, 1.223~GeV/c, 0.663),\\
($p_{\gamma,b2}^*,p^*_{\Upsilon,b2},f_{0,b2})=$(5.523~GeV/c, 1.271~GeV/c, 0.663) and \\
($p_{\gamma,b3}^*,p^*_{\Upsilon,b3},f_{0,b3})=$(5.526~GeV/c, 1.277~GeV/c, 0.663). As a result,
we get from Eq. (34):
\begin{eqnarray}
Br[P^+_{b1} \to {\gamma}p]=1.73\cdot10^{-4}Br[P^+_{b1} \to {\Upsilon(1S)}p],\nonumber\\
Br[P^+_{b2} \to {\gamma}p]=1.67\cdot10^{-4}Br[P^+_{b2} \to {\Upsilon(1S)}p],\nonumber\\
Br[P^+_{b3} \to {\gamma}p]=1.67\cdot10^{-4}Br[P^+_{b3} \to {\Upsilon(1S)}p].
\end{eqnarray}
The free total cross sections
$\sigma_{{\gamma}p \to P^+_{bi}\to {\Upsilon(1S)}p}(\sqrt{s},\Gamma_{bi})$ and
$\sigma_{{\gamma}n \to P^0_{bi}\to {\Upsilon(1S)}n}(\sqrt{s},\Gamma_{bi})$
for resonant $\Upsilon(1S)$ production in the
two-step processes (25)--(28) can be represented in the following forms [1, 4]:
\begin{equation}
\sigma_{{\gamma}p \to P^+_{bi}\to {\Upsilon(1S)}p}(\sqrt{s},\Gamma_{bi})=
\sigma_{{\gamma}p \to P^+_{bi}}(\sqrt{s},\Gamma_{bi})\theta[\sqrt{s}-(m_{\Upsilon(1S)}+m_N)]
Br[P^+_{bi} \to {\Upsilon(1S)}p],
\end{equation}
\begin{equation}
\sigma_{{\gamma}n \to P^0_{bi}\to {\Upsilon(1S)}n}(\sqrt{s},\Gamma_{bi})=
\sigma_{{\gamma}n \to P^0_{bi}}(\sqrt{s},\Gamma_{bi})\theta[\sqrt{s}-(m_{\Upsilon(1S)}+m_N)]
Br[P^0_{bi} \to {\Upsilon(1S)}n].
\end{equation}
Here, $\theta(x)$ is the usual step function.
According to Eqs.~(33), (34) and (38) these cross sections are equal to each other and they are proportional to $Br^2[P^+_{bi} \to {\Upsilon(1S)}p]$ and $Br^2[P^0_{bi} \to {\Upsilon(1S)}n]$, respectively.

According to [1, 4], we obtain the following expression for the total cross section for $\Upsilon(1S)$
production in ${\gamma}A$ interactions from the chains (25)--(28):
\begin{equation}
\sigma_{{\gamma}A\to {\Upsilon(1S)}X}^{({\rm sec})}(E_{\gamma})=
\sum_{i=1}^3\left[\sigma_{{\gamma}A\to P^+_{bi}\to{\Upsilon(1S)}p}^{({\rm sec})}(E_{\gamma})+
\sigma_{{\gamma}A\to P^0_{bi}\to{\Upsilon(1S)}n}^{({\rm sec})}(E_{\gamma})\right],
\end{equation}
where
\begin{eqnarray}
\sigma_{{\gamma}A\to P^+_{bi}\to{\Upsilon(1S)}p}^{({\rm sec})}(E_{\gamma})=\left(\frac{Z}{A}\right)
I_{V}[A,\sigma^{\rm eff}_{P_{bi}N}]\left<\sigma_{{\gamma}p \to P^+_{bi}}(E_{\gamma})\right>_A
Br[P^+_{bi} \to {\Upsilon(1S)}p],\nonumber\\
\sigma_{{\gamma}A\to P^0_{bi}\to{\Upsilon(1S)}n}^{({\rm sec})}(E_{\gamma})=\left(\frac{N}{A}\right)
I_{V}[A,\sigma^{\rm eff}_{P_{bi}N}]\left<\sigma_{{\gamma}n \to P^0_{bi}}(E_{\gamma})\right>_A
Br[P^0_{bi} \to {\Upsilon(1S)}n]
\end{eqnarray}
and
\begin{equation}
\left<\sigma_{{\gamma}n \to P^0_{bi}}(E_{\gamma})\right>_A=
\left<\sigma_{{\gamma}p \to P^+_{bi}}(E_{\gamma})\right>_A
\end{equation}
$$
=\int\int
P_A({\bf p}_t,E)d{\bf p}_tdE
\sigma_{{\gamma}p \to P^+_{bi}}(\sqrt{s_{\Upsilon(1S)}},\Gamma_{\rm med}^{bi})
\theta[\sqrt{s_{\Upsilon(1S)}}-(m_{\Upsilon(1S)}+m_N)].
$$
Here, $\sigma_{{\gamma}p \to P^+_{bi}}(\sqrt{s_{\Upsilon(1S)}},\Gamma_{\rm med}^{bi})$
is the "in-medium" cross section
for the $P^+_{bi}$ resonance production in ${\gamma}p$ collisions (25), $Z$ and $N$ are the numbers of
protons and neutrons in the target nucleus.
As above in Eq. (29), we assume that this cross section is equivalent to the free
cross section of Eq. (33) in which the vacuum decay width $\Gamma_{bi}$
is replaced by the in-medium width $\Gamma_{\rm med}^{bi}$ as given by Eqs.~(30)--(32) and vacuum
center-of-mass energy squared $s$, presented by the formula (8), is replaced by the in-medium expression (6).
The quantity $I_{V}[A,\sigma^{\rm eff}_{P_{bi}N}]$ in Eq. (43) is defined above by Eq. (4) in which one
needs to make the substitution $\sigma \to \sigma^{\rm eff}_{P_{bi}N}$.
Here, $\sigma^{\rm eff}_{P_{bi}N}$ is the $P^+_{bi}$, $P^0_{bi}$--nucleon effective absorption cross section.
This cross section can be represented [1, 4] as a sum of the inelastic cross section $\sigma_{P_{bi}N}$,
introduced above, and the additional to this cross section
effective $P^+_{bi}$, $P^0_{bi}$ absorption cross section associated with their decays in the nucleus.
From the standpoint of generality, we assume that the cross section $\sigma^{\rm eff}_{P_{bi}N}$
has the same value of 37 mb as was adopted in Ref. [1] for the $P_{ci}^+$--nucleon effective
absorption cross section $\sigma^{\rm eff}_{P_{ci}N}$.
\begin{figure}[htb]
\begin{center}
\includegraphics[width=16.0cm]{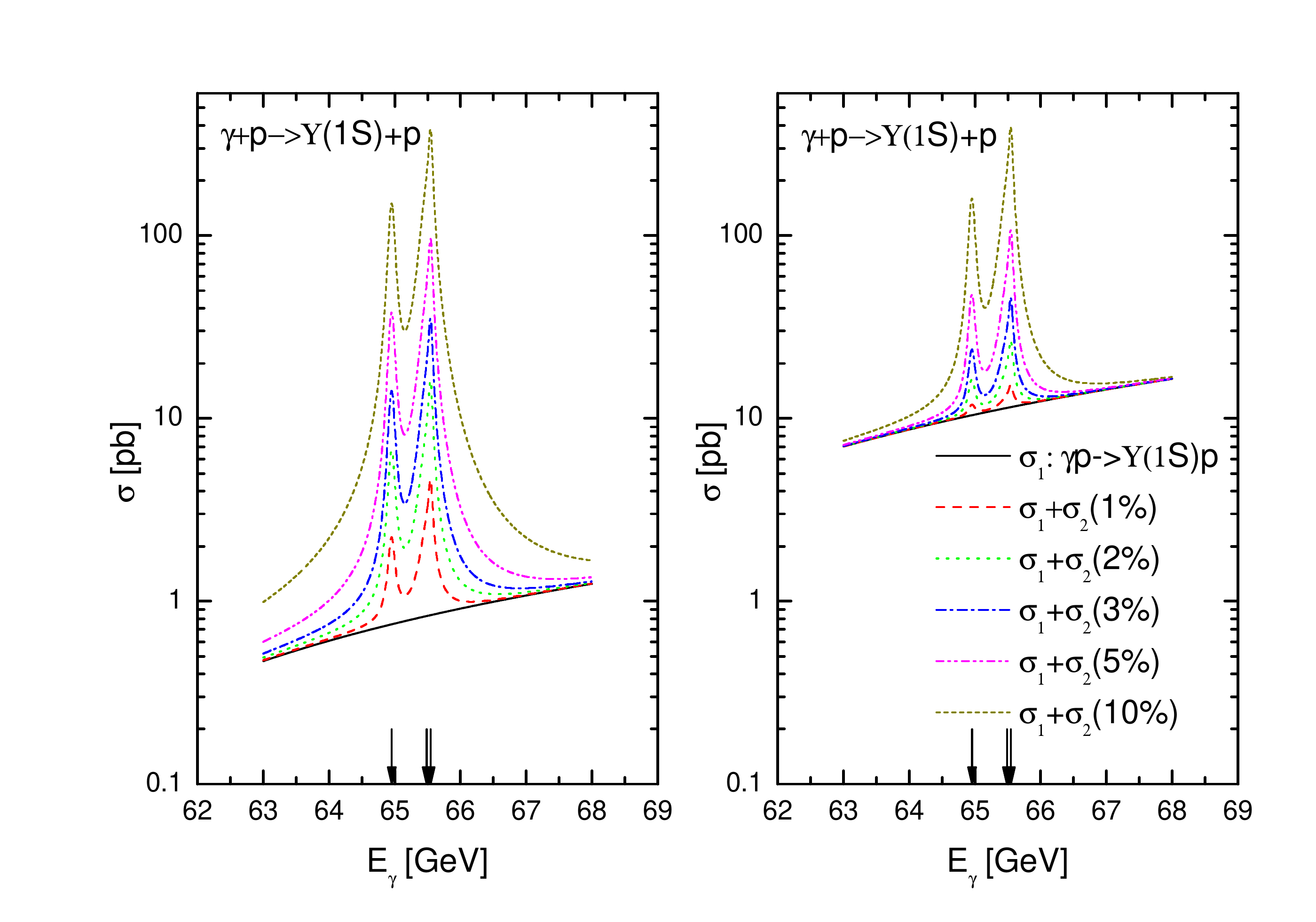}
\vspace*{-2mm} \caption{(Color online) The non-resonant total cross section for the reaction
${\gamma}p \to {\Upsilon(1S)}p$ (solid curves), calculated on the basis of Eqs. (22) (left panel)
and (10)--(20) (right panel). Incoherent sum of it and the total cross section for the
resonant $\Upsilon(1S)$ production in the processes ${\gamma}p \to P^+_{bi} \to {\Upsilon(1S)}p$,
($i=$1, 2, 3), calculated assuming that the resonances $P^+_{b1}$, $P^+_{b2}$ and $P^+_{b3}$
with the spin-parity quantum numbers $J^P=(1/2)^-$, $J^P=(1/2)^-$ and $J^P=(3/2)^-$
decay to ${\Upsilon(1S)}p$ with the lower allowed relative orbital angular momentum $L=0$
with all three branching fractions $Br[P^+_{bi} \to {\Upsilon(1S)}p]=1$, 2, 3, 5 and 10\%
(respectively, dashed, dotted, dashed-dotted, dashed-dotted-dotted and short-dashed curves),
as functions of laboratory photon energy $E_{\gamma}$.
The three arrows indicate the resonant energies $E^{\rm R1}_{\gamma}=64.952$ GeV,
$E^{\rm R2}_{\gamma}=65.484$ GeV and $E^{\rm R3}_{\gamma}=65.544$ GeV.}
\label{void}
\end{center}
\end{figure}
\begin{figure}[!h]
\begin{center}
\includegraphics[width=16.0cm]{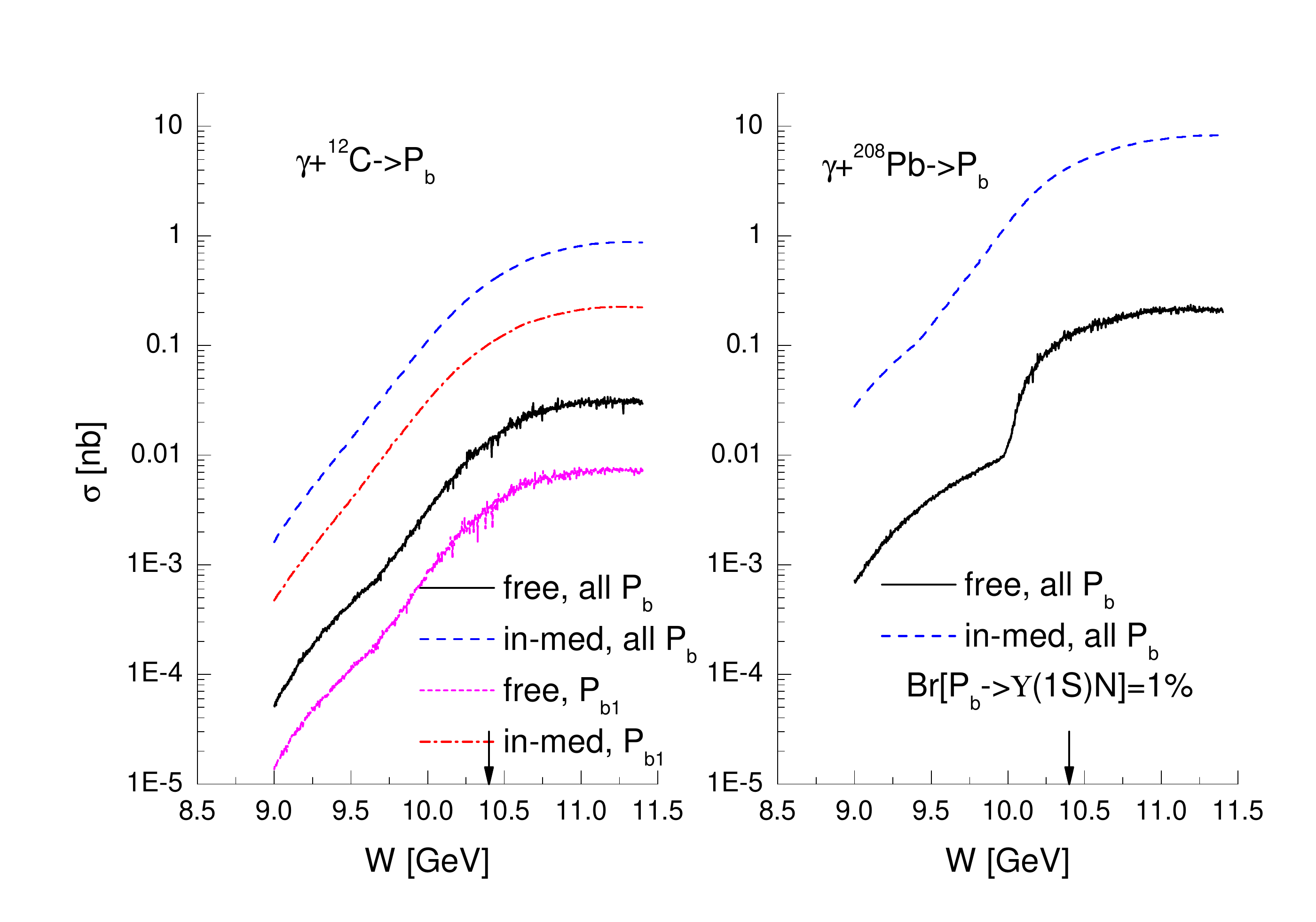}
\vspace*{-2mm} \caption{(Color online) Excitation functions for resonant production of
$P^{+}_{bi}$ and $P^{0}_{bi}$ ($i=$1, 2, 3) states
off $^{12}$C and $^{208}$Pb from the processes ${\gamma}p \to P^+_{bi}$
and ${\gamma}n \to P^0_{bi}$,
respectively, going on off-shell target nucleons, calculated
for $Br[P^+_{bi} \to {\Upsilon(1S)}p]=Br[P^0_{bi} \to {\Upsilon(1S)}n]=1$\% for all $i$ adopting free
(solid curves) and in-medium (dashed curves) $P^+_{bi}$,  $P^0_{bi}$ spectral functions.
$^{12}$C case: the same as above, but only for the processes ${\gamma}p \to P^+_{b1}$ and
${\gamma}n \to P^0_{b1}$, employing free (short-dashed) and in-medium (dotted-dashed)
$P^+_{b1}$ and $P^0_{b1}$ spectral functions.
The arrows indicate the threshold center-of-mass
energy for direct $\Upsilon(1S)$ photoproduction on a free target nucleon being at rest.}
\label{void}
\end{center}
\end{figure}
\begin{figure}[!h]
\begin{center}
\includegraphics[width=16.0cm]{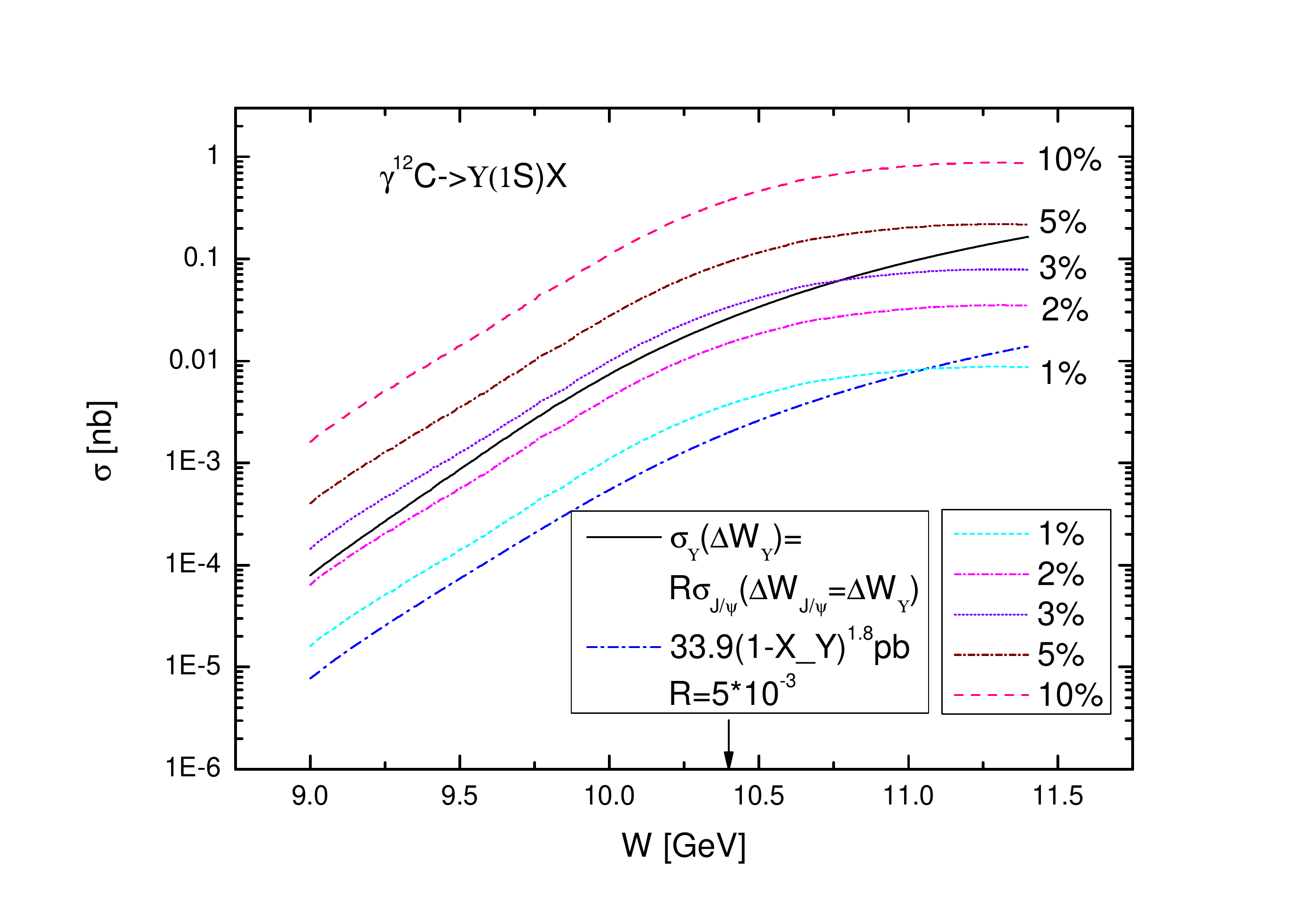}
\vspace*{-2mm} \caption{(Color online) Excitation functions for the non-resonant and resonant production
of $\Upsilon(1S)$ mesons off $^{12}$C from direct ${\gamma}N \to {\Upsilon(1S)}N$ and resonant
${\gamma}p \to P^+_{bi} \to {\Upsilon(1S)}p$ and ${\gamma}n \to P^0_{bi} \to {\Upsilon(1S)}n$ ($i=$1, 2, 3)
reactions going on off-shell target nucleons. The curves (solid and dotted-dashed),
corresponding to the non-resonant production of $\Upsilon(1S)$ mesons, are calculations by (3)
with elementary cross section $\sigma_{{\gamma}p \to {\Upsilon(1S)}p}$ in the forms of (10)--(20) and (22), respectively. The curves, belonging to their resonant production, are
calculations by (42) for branching ratios
$Br[P^+_{bi} \to {\Upsilon(1S)}p]=Br[P^0_{bi} \to {\Upsilon(1S)}n]=1$, 2, 3, 5 and 10\% for all $i$
adopting in-medium $P^+_{bi}$, $P^0_{bi}$ spectral functions. The arrow indicates the threshold
center-of-mass energy for direct $\Upsilon(1S)$ photoproduction on a free target nucleon being at rest.}
\label{void}
\end{center}
\end{figure}

\section*{3. Numerical results and discussion}

  The free elementary non-resonant $\Upsilon(1S)$ production cross section in the reaction
${\gamma}p \to {\Upsilon(1S)}p$, determined on the basis of Eqs. (22) (left panel)
and (10)--(20) (right panel), and the combined (non-resonant plus resonant (40))
total cross sections are presented in Fig. 3. From this
figure, one can see that the $P^+_b(11080)$ state appears as clear narrow independent peak at
$E_{\gamma}=$ 64.95 GeV in the combined cross section, while the $P^+_b(11125)$ and $P^+_b(11130)$
resonances exhibit itself here, due to low distance between their centroids (60 MeV),
as one distinct wide peak at $E_{\gamma}$ $\approx$ 65.50 GeV for two adopted choices (10)--(20)
and (22) for the background contribution, if $Br[P^+_{bi} \to {\Upsilon(1S)}p]=2$, 3, 5 and 10\%
($i=$1, 2, 3). In these cases, at laboratory photon energies around the peak energies the resonant
contributions are much larger than the non-resonant ones. Therefore, the background reaction will
not influence the direct observation of the hidden-bottom pentaquark production at these energies.
The peak values of the combined cross section reach tens and hundreds of picobarns,
if $Br[P^+_{bi} \to {\Upsilon(1S)}p]=2$ and 10\%, respectively
\footnote{$^)$It should be pointed out that the peak strengths of the combined cross section
of the reaction ${\gamma}p \to {\Upsilon(1S)}p$, corresponding to the
$P^+_b(11080)$ and $P^+_b(11125)$ states and obtained within the dipole Pomeron model in Ref. [21],
are about of 3 and 8 nb, respectively. These are much larger than those determined in the present work.}$^)$
.
But, they are much smaller than those of a few nanobarns for the reaction
${\gamma}p \to {J/\psi}p$ with $P^+_{ci}$ production [1]. This requires both the very high luminosities,
which will be accessible at future facilities such as proposed
electron--ion colliders EIC [22--24] and EicC [25, 26] in the U.S. and China, and large-acceptance detectors.
The strengths of these two peaks, obtained for
$Br[P^+_{bi} \to {\Upsilon(1S)}p]=$1\%, decrease essentially compared to the above cases and have
a peak values of about 2, 5 pb and 12, 15 pb for background contribution in the form of (22) and
(10)--(20), respectively. But, in former case, the $P_b^+$ signal to background ratio is larger
than that in the latter case by about of one order of magnitude. Therefore, it is natural to expect
that this signal can be distinguished from the background reaction as well, if it has the cross
section of about 1 pb in the energy region around energies $E_{\gamma}=$64.95 and $E_{\gamma}=$65.50 GeV.
To see experimentally
such two-peak structure in the combined total cross section of the reaction ${\gamma}p \to {\Upsilon(1S)}p$
\footnote{$^)$The $\Upsilon(1S)$ mesons could be identified via the muonic decays $\Upsilon(1S) \to {\mu^+}{\mu^-}$
with a brancing ratio of 2.48\% [34].}$^)$
,
it is enough to have the photon energy resolution and energy binning of the order of 20--30 MeV
\footnote{$^)$It should be noticed that, for example, in the GlueX experiment [5]
the $E_{\gamma}$ resolution was 20 MeV for a 10 GeV photon.}$^)$
.
Thus, the c.m. energy ranges
$M_{bi}-{\Gamma_{bi}}/2 < \sqrt{s} < M_{bi}+{\Gamma_{bi}}/2$ ($i=$1, 2) correspond to laboratory
photon energy regions of 64.894 GeV $< E_{\gamma} <$ 65.010 GeV
and 65.362 GeV $< E_{\gamma} <$ 65.607 GeV, i.e. ${\Delta}E_{\gamma}=$116 and 245 MeV for
$P^+_b(11080)$ and $P^+_b(11125)$, respectively. This means that to resolve the two
peaks in Fig. 3 the photon energy resolution and the energy bin size of the order of 20--30 MeV
are required. Finally, it is worth noting that the measurement of elastic bottomonium production on proton
close to threshold at electron-ion colliders will allow to determine the contribution of the so-called
trace anomaly term to the proton mass as well [50]. It has not been determined yet experimentally, nor by a lattice
QCD calculations [50]. The determination of this contribution would enable us to better understand the origin
of the total mass of the nucleon in terms of its constituents (quarks and gluons).
And in addition, it should be mentioned that the use of bottomonium
production at large W should allow one to shed light also on the contribution of the total
gluon angular momentum to the proton spin [50].

    Figure 4 shows the energy dependences of the total $P^+_{b1}$, $P^+_{b2}$,
$P^+_{b3}$ and $P^0_{b1}$, $P^0_{b2}$, $P^0_{b3}$
production cross section in ${\gamma}$$^{12}$C and $\gamma$$^{208}$Pb reactions as well as of
the total $P^+_{b1}$, $P^0_{b1}$ creation in ${\gamma}$$^{12}$C collisions.
They are calculated on the basis of Eqs.~(42), (43)
\footnote{$^)$By assuming that in Eq. (43) $Br[P^+_{bi} \to {\Upsilon(1S)}p]=Br[P^0_{bi} \to {\Upsilon(1S)}n]=1$
for all $i$ considered.}$^)$
in the scenarios with free and in-medium $P^+_{bi}$ and $P^0_{bi}$ spectral
functions for branching ratios $Br[P^+_{bi} \to {\Upsilon(1S)}p]=Br[P^0_{bi} \to {\Upsilon(1S)}n]=1$\%
($i=1$, 2, 3). It is seen that the hidden-bottom pentaquark
resonance formation is smeared out by Fermi motion of intranuclear nucleons.
It is a substantially enhanced for the in-medium case at all photon c.m.s. energies considered.
As is also easy to see, the contribution to the $\Upsilon(1S)$ production on nuclei, which will come from the intermediate $P^+_b(11080)$  and $P^0_b(11080)$ states, amounts approximately 25\% both
at subthreshold incident energies ($W < 10.4$ GeV) and at
above threshold beam energies ($W > 10.4$ GeV).
\begin{figure}[!h]
\begin{center}
\includegraphics[width=16.0cm]{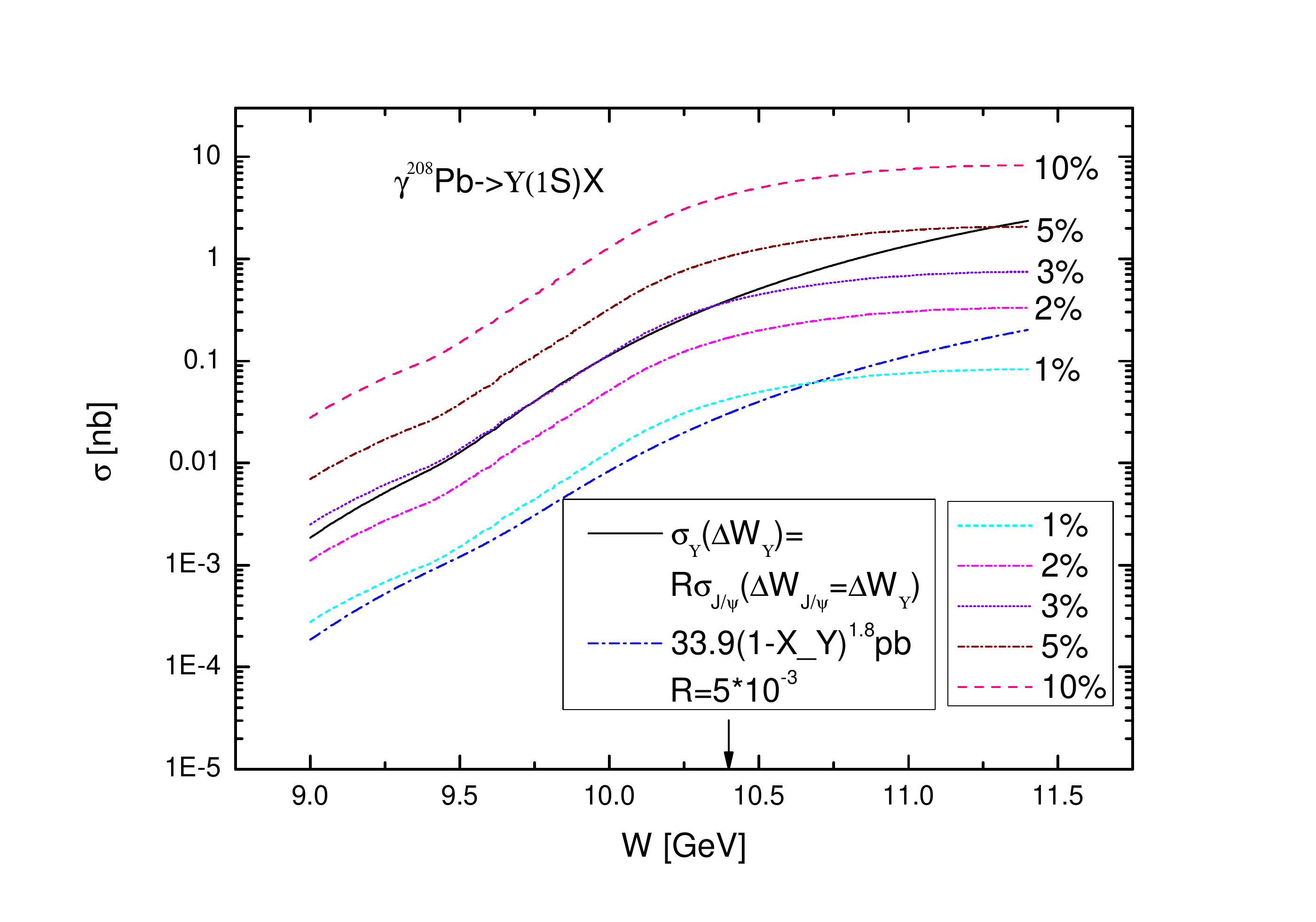}
\vspace*{-2mm} \caption{(Color online) The same as in figure 5, but for the $^{208}$Pb target nucleus.}
\label{void}
\end{center}
\end{figure}

    Excitation functions for non-resonant production of $\Upsilon(1S)$ mesons as well as for
their resonant production via $P^+_{b1}$, $P^+_{b2}$, $P^+_{b3}$ and
$P^0_{b1}$, $P^0_{b2}$, $P^0_{b3}$ resonances formation and decay in
${\gamma}$$^{12}$C and $\gamma$$^{208}$Pb collisions are given in Figs.~(5) and (6), respectively.
The former ones are calculated using Eq.~(3) for two employed options (10)--(20) and (22) for the
non-resonant elementary cross section $\sigma_{{\gamma}p \to {\Upsilon(1S)}p}$, whereas the latter ones
are determined using Eqs.~(42), (43) in the in-medium $P^+_{bi}$ and $P^0_{bi}$
spectral functions scenario and assuming that for all $i$ branching ratios
$Br[P^+_{bi} \to {\Upsilon(1S)}p]=Br[P^0_{bi} \to {\Upsilon(1S)}n]=1$, 2, 3, 5 and 10\%.
One can see that the non-resonant $\Upsilon(1S)$ yield and that from the production and decay of the
intermediate $P_{bi}^+$ and $P_{bi}^0$ resonances are comparable for both considered target nuclei if
$Br[P^+_{bi} \to {\Upsilon(1S)}p]=Br[P^0_{bi} \to {\Upsilon(1S)}n]=1$ and 3\% when the background cross
section $\sigma_{{\gamma}p \to {\Upsilon(1S)}p}$ is used in the forms of (22) and (10)--(20), respectively.
But, if these branching ratios are more than 1 and 3\%, correspondingly, the resonant $\Upsilon(1S)$
production cross section is much larger, especially at subthreshold beam energies, than the non-resonant one
and their relative strength is governed by the ratios.

  Thus, the presence of the $P_{bi}^+$ and $P_{bi}^0$
pentaquark states leads to additional (and essential) enhancement in the behavior of the total $\Upsilon(1S)$
production cross section on nuclei both below and above threshold and the strength of this enhancement
is strongly determined by the branching fractions of their decays to $\Upsilon(1S)p$ and $\Upsilon(1S)n$
final states, respectively. These fractions can be accurately studied experimentally at electron-ion
colliders also via the bottomonium excitation function measurements on nuclear targets near threshold
and comparison their results with the calculations on the basis of the present model
with known total cross sections of direct processes (1) and (2)
\footnote{$^)$If these cross sections are different, then in Eq. (3) one needs to perform the following
substitution $\left<\sigma_{{\gamma}p \to {\Upsilon(1S)}p}(E_{\gamma})\right>_A$ $\to$
$(Z/A)\left<\sigma_{{\gamma}p \to {\Upsilon(1S)}p}(E_{\gamma})\right>_A+(N/A)
\left<\sigma_{{\gamma}n \to {\Upsilon(1S)}n}(E_{\gamma})\right>_A$.}$^)$
.
The collected statistics in these measurements, especially on heavy target nuclei and at above
threshold energies where the resonant $\Upsilon(1S)$ production cross section reaches
the values $\sim$ 1--10 nb for above branching fractions $\sim$ 5--10\%,
is expected to be substantially higher than that, which could be achieved
in measurements on the nucleon target (cf. Figs. 5, 6 and 3). This should enable a
more accurate determination of these fractions in the measurements on nuclear targets.

\section*{4. Summary}

 In this work we have calculated the absolute excitation functions for the non-resonant and resonant
photoproduction of $\Upsilon(1S)$ mesons off protons at threshold incident photon laboratory energies of
63--68 GeV by accounting for direct (${\gamma}p \to {\Upsilon(1S)}p$) and two-step
   (${\gamma}p \to P^+_b(11080) \to {\Upsilon(1S)}p$,
   ${\gamma}p \to P^+_b(11125) \to {\Upsilon(1S)}p$,
   ${\gamma}p \to P^+_b(11130) \to {\Upsilon(1S)}p$) $\Upsilon(1S)$ production channels
within different scenarios for the non-resonant total cross section of elementary reaction ${\gamma}p \to {\Upsilon(1S)}p$ and for branching ratios of the decays
   $P^+_b(11080) \to {\Upsilon(1S)}p$, $P^+_b(11125) \to {\Upsilon(1S)}p$, $P^+_b(11130) \to {\Upsilon(1S)}p$.
Also, an analogous functions for photoproduction of $\Upsilon(1S)$ mesons on $^{12}$C and $^{208}$Pb
target nuclei in the near-threshold center-of-mass beam energy region of 9.0--11.4 GeV have been
calculated by considering incoherent direct (${\gamma}N \to {\Upsilon(1S)}N$) and two-step
   (${\gamma}p \to P^+_b(11080) \to {\Upsilon(1S)}p$,
   ${\gamma}p \to P^+_b(11125) \to {\Upsilon(1S)}p$,
   ${\gamma}p \to P^+_b(11130) \to {\Upsilon(1S)}p$ and
   ${\gamma}n \to P^0_b(11080) \to {\Upsilon(1S)}n$,
   ${\gamma}n \to P^0_b(11125) \to {\Upsilon(1S)}n$,
   ${\gamma}n \to P^0_b(11130) \to {\Upsilon(1S)}n$) $\Upsilon(1S)$ production processes
within a nuclear spectral function approach.
It was shown that the $P^+_b(11080)$ state appears as clear narrow independent peak at
$E_{\gamma}=$ 64.95 GeV in the combined (non-resonant plus resonant) cross section on proton target,
while the $P^+_b(11125)$ and $P^+_b(11130)$ resonances exhibit itself here,
due to low distance between their centroids (60 MeV),
as one distinct wide peak at $E_{\gamma}$ $\approx$ 65.50 GeV for two adopted options
for the background contribution, if $Br[P^+_{bi} \to {\Upsilon(1S)}p]=2$, 3, 5 and 10\%
($i=$1, 2, 3). The peak values of the combined cross section reach tens and hundreds of picobarns,
if $Br[P^+_{bi} \to {\Upsilon(1S)}p]=2$ and 10\%, respectively. Therefore, a detailed scan of the
$\Upsilon(1S)$ total photoproduction cross section on a proton target in the near-threshold energy
region in future high-precision experiments at electron-ion colliders should give a definite
result for or against the existence of the non-strange hidden-bottom pentaquark states and
clarify their decay rates.

It was also demonstrated that the presence of the $P_{bi}^+$ and $P_{bi}^0$ pentaquark states in
$\Upsilon(1S)$ photoproduction on nuclei
leads to additional (and essential) enhancement in the behavior of the total $\Upsilon(1S)$
production cross section on nuclei both below and above threshold and the strength of this enhancement
is strongly determined by the branching fractions of their decays to $\Upsilon(1S)p$ and $\Upsilon(1S)n$
final states, respectively. This offers an indirect possibility of studying of these fractions experimentally
at the future high-luminosity electron-ion colliders EIC and EicC in the U.S. and China
also via the near-threshold bottomonium excitation function measurements on nuclear targets.
The collected statistics in these measurements, especially on heavy target nuclei and at above
threshold energies where the resonant $\Upsilon(1S)$ production cross section reaches
the values $\sim$ 1--10 nb for above branching fractions $\sim$ 5--10\%, is expected to be substantially
higher than that, which could be achieved in measurements on the nucleon target. This should enable a
more accurate determination of these fractions in the measurements on nuclear targets.

\end{document}